\documentclass[modern]{aastex631}

\newcommand{\teff}{$T_{\mathrm{eff}}$}

\newcommand{\numax}{$\nu_{\mathrm{max}}$}
\newcommand{\dnu}{$\Delta\nu$}

\newcommand{\kepler}{\textit{Kepler}}

\begin{document}

\title{Extension of the Asfgrid for correcting asteroseismic large frequency separations}

\author[0000-0002-4879-3519]{Dennis Stello}
\affiliation{School of Physics, University of New South Wales, NSW 2052, Australia}
\affiliation{Sydney Institute for Astronomy (SIfA), School of Physics, University of Sydney, NSW 2006, Australia}
\affiliation{Stellar Astrophysics Centre, Department of Physics and Astronomy, Aarhus University, DK-8000 Aarhus C, Denmark}
\affiliation{ARC Centre of Excellence for All Sky Astrophysics in Three Dimensions (ASTRO-3D)}

\author[0000-0002-0920-809X]{Sanjib Sharma}
\affiliation{Sydney Institute for Astronomy (SIfA), School of Physics, University of Sydney, NSW 2006, Australia}
\affiliation{ARC Centre of Excellence for All Sky Astrophysics in Three Dimensions (ASTRO-3D)}




\begin{abstract}

The asteroseismic scaling relation, \dnu\ $\approx\rho^{0.5}$, linking a star's large frequency separation, \dnu, and its mean density, $\rho$, is not exact. Yet, it provides a very useful way to obtain fundamental stellar properties. Common ways to make the relation more accurate is to apply correction factors to it. Because the corrections depend on stellar properties, such as mass, \teff, and metallicity, it is customary to interpolate these properties over stellar model grids that include both \dnu, measured from adiabatic frequencies of the models, and the models' stellar density; hence linking both sides of the scaling relation. A grid and interpolation tool widely used for this purpose, known as Asfgrid, was published by \citet{SharmaStello16}. Here, we present a significant extension of Asfgrid to cover higher- and lower-mass stars and to increase the density of grid points, especially in the low-metallicity regime.

\end{abstract}

\keywords{stars: fundamental parameters --- stars: oscillations --- stars: interiors}


\section{} 
\begin{figure}
\plotone{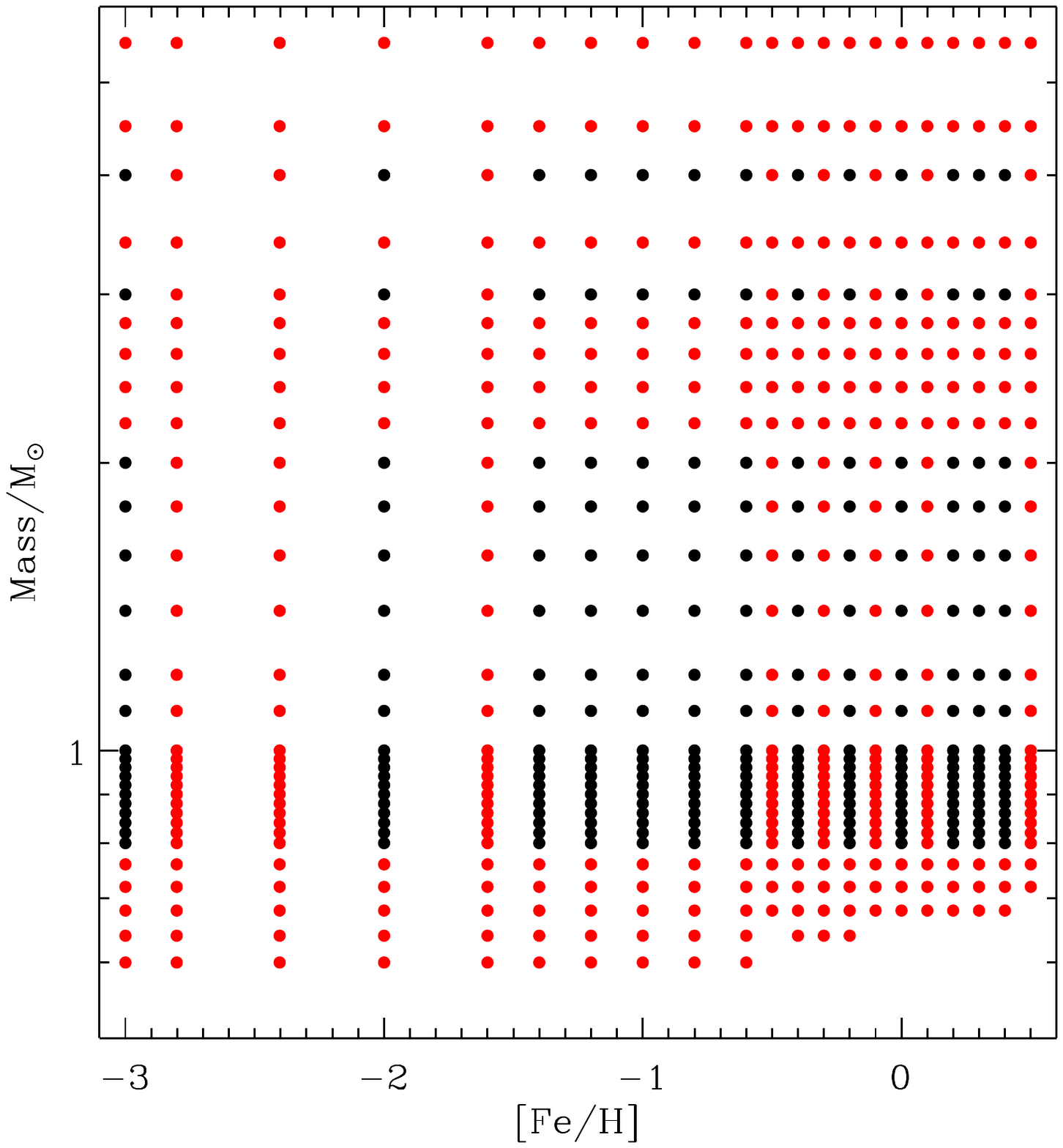}
\caption{Grid points in Mass and [Fe/H] of the original grid from \citet{Sharma16} (black) and our new extension (red).  \label{fig:gridpoints}}
\end{figure}

The initial release of the Asfgrid by \citet{SharmaStello16} was published as part of an investigation of the \kepler\ red giant sample \citep{Sharma16}. It was therefore made to cover the typical ranges in mass and [Fe/H] of those stars with most grid points concentrated in the parameter range where most \kepler\ giants were found. Because the community is increasingly studying oscillations in stars in the more extreme ends of the mass-[Fe/H] parameter space we have extended the Asfgrid from its initial 247 grid points in mass-[Fe/H] (see black points in Figure~\ref{fig:gridpoints}) to now having 632 points (black and red points). This extension was mainly made to cover the low-mass stars ($0.6 < M/\mathrm{M}_\odot< 0.8$) and to ensure that interpolations become more accurate at low metallicities ($-3 < \mathrm{[Fe/H]} < -1.4$) where the initial grid was sparse. In addition, we have extended the grid to slightly higher mass ($M/\mathrm{M}_\odot \leq 5.5$) and higher metallicity ([Fe/H] $\leq 0.5$) relative to the initial release.

For the extension we used identical input physics and the same version of MESA (v6950) \citep{Paxton11,Paxton13,Paxton15} to calculate the stellar models ensuring consistency with the initial grid, which forms part of the final grid. The evolution was terminated at the tip of the asymptotic giant branch. We also used the same version of GYRE (v3.1) \citep{TownsendTeitler13} with the same settings to calculate the adiabatic frequencies and the \citet{White11} method to derive \dnu\ from those frequencies. Further details about the models are given in \citet[][section 2.3]{Sharma16}.

Like the initial release, the two main things that Asfgrid can estimate are: 
\begin{itemize}

    \item {\bf Asteroseismic frequencies \numax\ and \dnu\ and the correction factor $f_{\Delta \nu}$}: These are estimated from surface gravity ($\log g$), effective temperature (\teff), metallicity ($Z$), and mass ($M$). The interpolation is done over the model grid in $Z$, $M$, evolution state ($E_{\rm state}$), and $\gamma$, where  
    \begin{equation}
        \gamma=\log(T_{\rm eff})+ 0.01 \log g  \left[\tanh\left(\frac{\log g-4.5}{0.25}\right)+1\right] \frac{1}{2}. 
    \end{equation}
    We introduce $\gamma$ because we ideally want to interpolate the stellar properties over age. However, age is not an observable, while  $T_{\rm eff}$ and $\log g$, which are observables, are not monotonic with age. Hence, $\gamma$ is designed to be roughly monotonic with age.  The correction factor, $f_{\Delta \nu}$, is
    estimated alongside using the relation
    \begin{eqnarray}
    f_{\Delta \nu} &=& \frac{\Delta \nu}{135.1 \ \mu{\rm Hz}}\left(\frac{\rho}{\rho_{\odot}}\right)^{-1/2} \\
    &=& \frac{\Delta \nu}{135.1 \ \mu{\rm Hz}} \left(\frac{g}{g_{\odot}}\right)^{-3/4}\left(\frac{M}{{\rm M}_{\odot}}\right)^{1/4}
    \end{eqnarray}

    \item {\bf Stellar mass $M$ and radius $R$}: They are estimated from \teff, $Z$, \numax, and \dnu. The interpolation is done over the grid in $Z$, $M_{\rm seismic}$, $E_{\rm state}$, and $\gamma$, where  
    \begin{eqnarray}
        M_{\rm seismic} &= & \left(\frac{g}{g_{\odot}}\right)^3 \left(\frac{\Delta \nu}{\Delta \nu_{\odot}}\right)^{-4} {\rm M}_{\odot}\ {\rm with}\\
        g &=& g_{\odot} \left(\frac{\nu_{\rm max}}{\nu_{{\rm max},\odot}}\right) \left(\frac{T_{\rm eff}}{T_{{\rm eff},\odot}}\right)^{1/2}
    \end{eqnarray}
    The following solar reference values are used, $g_{\odot}=4.43796\ {\rm cm\ s^{-2}}$, $T_{{\rm eff}, \odot}=5777\ {\rm K}$,  $\Delta \nu_{\odot}=135.1\ \mu{\rm Hz}$, $\nu_{{\rm max},\odot}=3090\ \mu{\rm Hz}$. 

\end{itemize}

The new extended grid, along with the interpolation code, can be accessed at \url{http://www.physics.usyd.edu.au/k2gap/Asfgrid/}. 
If you use the extended Asfgrid (v0.0.6) for your research, we kindly ask you to cite both \citet{Sharma16} and this Research Note.



\bibliography{bib_complete}{}
\bibliographystyle{aasjournal}



\end{document}